\documentclass[a4paper,twoside]{article}

\usepackage{epsfig}
\usepackage{subfig}
\usepackage{calc}
\usepackage{amssymb}
\usepackage{amstext}
\usepackage{amsmath}
\usepackage{amsthm}
\usepackage{multicol}
\usepackage{pslatex}
\usepackage{apalike}
\usepackage{algorithm2e}
\usepackage[bottom]{footmisc}
\usepackage{SCITEPRESS}
\usepackage{tikz}
\usetikzlibrary{quantikz2}
\usepackage{tabularx}
\newcolumntype{C}{>{\centering\arraybackslash}X}
\usepackage[hidelinks]{hyperref}
\usepackage[capitalise]{cleveref}
\creflabelformat{equation}{#2#1#3}

\begin{document}

\title{Quantum Advantage Actor-Critic for Reinforcement Learning}

\author{\authorname{Michael Kölle\sup{1*}, Mohamad Hgog\sup{1*}, Fabian Ritz\sup{1}, Philipp Altmann\sup{1}, Maximilian Zorn\sup{1}, Jonas Stein\sup{1} and Claudia Linnhoff-Popien\sup{1}}
\affiliation{\sup{1}Institute of Informatics, LMU Munich, Munich, Germany}
\email{michael.koelle@ifi.lmu.de}
}

\abstract{
Quantum computing offers efficient encapsulation of high-dimensional states. In this work, we propose a novel quantum reinforcement learning approach that combines the Advantage Actor-Critic algorithm with variational quantum circuits by substituting parts of the classical components. This approach addresses reinforcement learning's scalability concerns while maintaining high performance. We empirically test multiple quantum Advantage Actor-Critic configurations with the well known Cart Pole environment to evaluate our approach in control tasks with continuous state spaces. Our results indicate that the hybrid strategy of using either a quantum actor or quantum critic with classical post-processing yields a substantial performance increase compared to pure classical and pure quantum variants with similar parameter counts. They further reveal the limits of current quantum approaches due to the hardware constraints of noisy intermediate-scale quantum computers, suggesting further research to scale hybrid approaches for larger and more complex control tasks.
}

\onecolumn \maketitle \normalsize \setcounter{footnote}{0} \vfill
\def\thefootnote{*}\footnotetext{Authors contributed equally to this work.}

\section{\uppercase{Introduction}} \label{sec:intro}
The field of quantum computing (QC) is commonly credited with the potential to solve highly complex problems with considerable speedup compared to classical computers, as it leverages the unique properties of quantum mechanics \cite{nielsen2010quantum}. This potential will be realized once reliable and usable quantum hardware becomes available, as discussed in \cite{Preskill_2018}. Still, even on QC hardware of current size, achievements in domains like quantum cryptography \cite{Pirandola_2020,Shor_1997}, quantum chemistry \cite{Cao_2019,bauer_quantum_2020,dral_quantum_2020}, and quantum optimization \cite{https://doi.org/10.48550/arxiv.1602.07674,Cerezo2020VariationalQA,farhi_quantum_2014} have shown promising application progress of QC to real-world problems. Inspired by the success of applying classical machine learning (ML) techniques to problems like image recognition, reinforcement learning, and natural language processing, the domain of quantum machine learning (QML) has also recently progressed through applying ML methods to and on quantum hardware. The common goal of QML algorithms is to leverage the inherent parallelism of quantum interference, potentially accelerating machine learning models' training and inference speed \cite{Biamonte_2017}. 

Reinforcement learning (RL) is a machine learning subfield focused on training agents to interact with an environment and learn from their experiences. RL has achieved remarkable success in applications such as game playing (e.g., AlphaGo \cite{silver_mastering_2017}), robotics \cite{doi:10.1177/0278364913495721}, and autonomous driving \cite{DBLP:journals/corr/YouPWL17}. However, the performance of classical RL algorithms is often limited by their sample inefficiency. This leads to slow convergence and requires a large number of interactions with the environment. Since quantum computing offers the efficient encapsulation of high-dimensional states as one of its significant benefits, recent research interest in quantum reinforcement learning (QRL) has emerged \cite{https://doi.org/10.48550/arxiv.2211.03464}.

In this paper, we focus on actor-critic methods \cite{NIPS1999_6449f44a}, a popular class of RL algorithms combining policy-based and value-based approaches. We specifically examine the Advantage Actor-Critic (A2C) algorithm, first proposed by Mnih et al., \cite{https://doi.org/10.48550/arxiv.1602.01783}. The A2C algorithm is well understood in the classical field and has been extensively benchmarked for control problems (e.g., \cite{andrychowicz2021matters}). In the younger context of QRL, research has been heavily focused on value-based methods like Deep Q-Networks \cite{mnih_human-level_2015}; however, this work focuses on actor-critic methods.

In this work, we explore variational quantum circuits (VQC) for policy gradient methods like the A2C algorithm. VQCs are parameterized quantum circuits trained using classical optimization techniques, suitable for the current era of noisy intermediate scale quantum computers \cite{Preskill_2018}. For our quantum variations of the A2C algorithm, we utilize VQCs, replacing classical neural networks for both the actor and critic. We evaluate and showcase the potential of VQCs to enhance the algorithm's learning efficiency and accuracy in solving simple control problems. We test our implementation on the well-understood but non-trivial \textit{Cart Pole} control environment provided by the OpenAI gymnasium of RL domains \cite{brockman2016openai}. This benchmark problem is chosen for its ability to highlight the potential benefits of the quantum A2C algorithm, such as faster convergence and improved stability, in solving control tasks with continuous state spaces.

We conducted experiments to evaluate different quantum, hybrid, and classical A2C configurations. First, we compared the performance of our quantum A2C configurations with the classical A2C algorithm, ensuring a similar number of parameters. Furthermore, we investigated the effectiveness of a VQC with a classical post-processing layer. Finally, we juxtaposed the performance of this hybrid approach with a classical A2C algorithm, maintaining a similar parameter count. Our findings indicate that the quantum hybrid A2C approach demonstrates a substantial performance advantage over the purely quantum and classical methods. However, it is essential to note that the quantum A2C algorithm faces potential limitations and challenges, such as scalability issues and hardware constraints related to the current era of NISQ computers. These limitations may impact the generalizability of our results, and further research is required to fully understand the performance of quantum A2C algorithms in more extensive and complex control tasks.

This work is structured as follows. First, we discuss the preliminaries in \cref{sec:preliminaries} and related works in \cref{sec:related-work}. We introduce our quantum A2C and hybrid quantum A2C architectures in \cref{sec:approach}. We then go into detail about our experimental setup in \cref{sec:experimental-setup} and the results in \cref{sec:results}. Lastly, we conclude with summarizing our findings and indicating future work in \cref{sec:conclusion}. 
\section{\uppercase{Preliminaries}} \label{sec:preliminaries}

\subsection{Policy Gradient and Actor-Critic Methods}

Policy gradient methods is a category within reinforcement learning, which directly optimizes policies by estimating optimal policy weights using gradient ascent, circumventing explicit trade-off requirements between exploration and exploitation and learning stochastic policies for state space exploration \cite{NIPS1999_464d828b}. Distinct from Q-learning's $\epsilon$-greedy strategy, they adeptly manage high-dimensional and continuous action spaces and elude perceptual aliasing by generating probability distributions \cite{https://doi.org/10.48550/arxiv.2007.02151}. Formally, policy gradient methods optimize policy parameters $\theta$ to amplify the expected cumulative reward, mathematically expressed as: 

\begin{equation}
J(\theta) = \mathbb{E}_{\pi_\theta} [R] 
\end{equation}
where $J(\theta)$ represents the objective function and $\pi_\theta$ signifies a policy. Utilizing Monte Carlo samples, the gradient of $J(\theta)$ relative to $\theta$ can be estimated and updated through stochastic gradient ascent, with the algorithm seeking optimal $\theta^*$.

Actor-critic methods synergize value-based and policy-based strategies, where the actor (policy function $\pi_\theta(s,a)$) and critic (value function $Q_w(s,a)$) learn their respective functionalities, often modeled using non-linear neural network function approximators in deep RL \cite{sutton2018reinforcement}. In essence: 
\begin{equation}
Q(s,a; w) \approx Q_\pi(s,a)
\end{equation}
with the actor updating policy parameters via gradients from the policy gradient theorem, and the critic estimating the value function for the current policy.

Various actor-critic algorithms, such as Asynchronous Advantage Actor-Critic (A3C), Deep Deterministic Policy Gradient, and Proximal Policy Optimization, employ the actor-critic framework to address challenges in continuous or substantial state spaces and stochastic policies \cite{https://doi.org/10.48550/arxiv.1602.01783,https://doi.org/10.48550/arxiv.1509.02971,https://doi.org/10.48550/arxiv.1707.06347}. Specifically, Advantage Actor-Critic (A2C) utilizes the advantage function:
\begin{equation}
    A(s, a) = Q(s, a) - V(s)
\end{equation}
to update policies based on actions' relative efficacy in given states, demonstrating enhanced accuracy and rapid convergence, thereby forming a robust focal point for further RL research.

\subsection{Variational Quantum Circuit}
\label{sec:vqc}
To comprehend the concept of variational quantum circuits, examining the underlying principles of quantum circuits is essential. Similar to classical circuits, they consist of individual components: quantum wires and quantum gates. Each quantum wire represents a quantum bit analogous to a classical circuit wire representing a classical bit. In contrast, quantum gates perform unitary transformations on the quantum bits, akin to how classical gates perform operations on classical bits \cite{homeister2018quantum}.

VQC, also known as a parametrized quantum circuit, is a quantum algorithm that employs a classical optimization process to prepare a quantum state encoding the solution to a specific problem. A VQC can be divided into different parts. In the first part, the classical input data enters the circuit as arguments and is encoded into quantum states using the initial quantum gates in the circuit. After encoding the data, qubits are entangled by controlled gates and rotated by parameterized rotation gates. This sequence of operations can be repeated multiple times with different parameters; each repetition is called a layer. The qubits are then measured, and the results are decoded into output information \cite{9144562}. The parameters in the layers are then updated with classical optimization algorithms, such as gradient descent or stochastic gradient descent, to optimize the algorithm's objective function. This optimization is performed step-by-step, with the parameters updated in each iteration \cite{pennylane_variational_circuit_2022}.

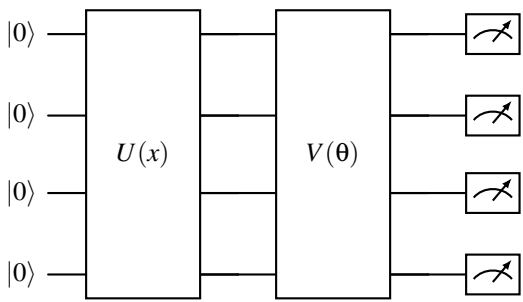
\begin{figure}[!htbp]
\begin{center}
\begin{quantikz}
\lstick{\ket{0}} & \gate[wires=4][1.5cm]{U(x)} & \qw & \gate[wires=4][1.5cm]{V(\theta)} & \qw & \meter{} \\
\lstick{\ket{0}} && \qw && \qw & \meter{} \\
\lstick{\ket{0}} && \qw && \qw & \meter{} \\
\lstick{\ket{0}} && \qw && \qw & \meter{}
\end{quantikz}
\end{center}
\captionsetup{labelfont=bf, format=plain}
\caption{The variational quantum circuit structure comprises three main parts: the $U(x)$-block represents the state preparation part, the $V(\theta)$-block constitutes the variational part containing trainable parameters $\theta$, and the quantum measurement layer.}
\label{fig:vqc}
\end{figure}

One of the notable advantages of such circuits is their robustness to quantum noise \cite{Chen_2022}, which can be beneficial for NISQ era quantum computers \cite{Preskill_2018}. Moreover, they have demonstrated significant success in various machine learning areas, such as classification \cite{Chen_2021} and natural language processing \cite{https://doi.org/10.48550/arxiv.2110.06510}.

\section{\uppercase{Related Work}} \label{sec:related-work}
Previous studies have explored various aspects of QRL, such as developing quantum algorithms and circuits for policy-based and value-based methods, exploring the potential advantages of QRL over classical RL, and implementing QRL in real-world problems \cite{KwakQRL_With_pennylane}.

This work focuses on applying VQC to RL methods, which is why in this chapter, we will discuss related works in this area, specifically highlighting the use of VQC.

One of the earliest works introducing the idea of using VQC for RL is presented in \cite{https://doi.org/10.48550/arxiv.1907.00397}. The authors demonstrated a proof-of-principle using a VQC to approximate the Q-value function in Deep Q-Learning (DQN) algorithm and tested the quantum version of the DQN on discrete environments. In work \cite{Skolik_2022}, the authors extended the previous achievements in DQN by exploring its performance in continuous environments. Further improvements in DQN, such as Double DQN, have also been implemented within the VQC architecture. For example, in the study \cite{https://doi.org/10.48550/arxiv.2202.12180}, the authors applied QRL to solve a robot navigation task using different VQC architectures with and without data re-uploading. In addition, in \cite{https://doi.org/10.48550/arxiv.2008.07524}, VQCs were used in both DQN and Double DQN algorithms in different continuous OpenAI Gym environments.

Moreover, significant progress has been made in developing policy-based methods frameworks for QRL that directly focus on learning policy functions \cite{https://doi.org/10.48550/arxiv.2212.09328}. For example, the works \cite{https://doi.org/10.48550/arxiv.2103.05577} \cite{https://doi.org/10.48550/arxiv.2203.10591} used a VQC for the basic Monte Carlo policy gradient algorithm REINFORCE.

In the field of actor-critic methods, various research approaches utilize QRL and implement VQCs as the actor and critic networks. For instance, the authors of \cite{https://doi.org/10.48550/arxiv.2108.06849} proposed a policy-VQC for the Proximal Policy Optimization algorithm, where the neural network of the actor was replaced with a VQC. Another study \cite{https://doi.org/10.48550/arxiv.2112.11921} introduced a soft actor-critic algorithm that employs VQCs as the policy function approximator. Notably, this study demonstrated that quantum models could perform better than classical models even with few parameters.
Furthermore, a recent work \cite{https://doi.org/10.48550/arxiv.2301.05096} proposed a quantum version of the A3C algorithm as a novel QRL approach. The author used a quantum A3C algorithm incorporating multiple VQC-based agents (both actor and critic) trained on different environments. The results showed that the proposed quantum algorithm outperformed traditional A3C in terms of convergence speed and sample efficiency.

During the NISQ era, quantum computers have limited qubits available, making it challenging to process large amounts of input data. A combination of VQC and neural networks was employed to overcome this challenge. For instance, in \cite{Mari_2020}, a pre-processing neural network was used to reduce the input dimension to match the number of qubits in the VQC, enabling efficient processing of image classification tasks. Similarly, in \cite{Chen_2021} \cite{Chen_2022}, tensor networks were used to reduce input dimensionality. Moreover, post-processing techniques such as neural network layers were used to reshape the VQC output in \cite{https://doi.org/10.48550/arxiv.2301.05096} and \cite{https://doi.org/10.48550/arxiv.2203.10591} or to enhance the VQC's expressive power for optimal results, as demonstrated in \cite{https://doi.org/10.48550/arxiv.2203.14348} and \cite{https://doi.org/10.48550/arxiv.2212.06663}.

Overall, the studies presented in this chapter showcase the potential of VQC in various RL methods, including DQN, policy gradient algorithms, and actor-critic methods. As a result, applying VQC to RL has become a promising avenue for future research. In this paper, we aim to build upon these advances by exploring the use of VQC in Actor-Critic methods and investigating their performance in different problem settings. We will assess the potential benefits and limitations of using VQC in RL problems and contribute to the ongoing development of quantum approaches in reinforcement learning.
\section{Approaches}
\label{sec:approach}

\subsection{Advantage Actor-Critic Algorithm}
\label{sec:a2c}

\begin{algorithm}[t]
\SetAlgoNlRelativeSize{-1}
\SetNlSty{}{}{}
\caption{Advantage Actor-Critic Algorithm \protect\cite{https://doi.org/10.48550/arxiv.1602.01783}}
\label{alg:a2c}
Initialize the environment\;
Initialize the actor-network $\pi_{\theta}(s)$\;
Initialize the critic-network $V_{\omega}(s)$\;
Initialize Adam optimizer for $\theta$ and $\omega$\;

\For{each episode}{
   Initialize next observation state $s_{t}$ $env.reset()$\;
   Initialize done $d = \text{False}$\;

   \For{each iteration $t$}{
      Select an action $a_t$ based on $\pi_{\theta}(s_t)$\;
      Execute action $a_t$ in the environment\;
      Observe reward $r_t$, new state $s_{t+1}$, and done $d$\;
      \If{$d = \text{True}$}{
         \textbf{break}\;
      }
      Calculate TD target value $y = r_t + \gamma V_{\omega}(s_{t+1})$\;
      Calculate advantage $A = y - V_{\omega}(s_t)$\;
      Update the actor-network by minimizing the loss $L_\text{actor}(\theta) = -A \log \pi_{\theta}(a_t)$\;
      Update the critic-network by minimizing the loss $L_\text{critic}(\omega) = (y - V_{\omega}(s_t))^2$\;
      Update the state $s_t = s_{t+1}$\;
   }
}
\end{algorithm}

The first step in the A2C algorithm (\cref{alg:a2c}) involves initializing both the actor-network $\pi_{\theta}(s)$ and the critic network $V_{\omega}(s)$ with random parameters $\theta$ and $\omega$. The actor-network maps the current state of the environment to an appropriate action, while the critic network maps the current state to its corresponding state value. Then, initialize the stochastic gradient descent optimizer Adam \cite{https://doi.org/10.48550/arxiv.1412.6980} to update the parameters $\theta$ and $ \omega$ throughout the optimization process.

Once the networks have been initialized, the algorithm enters a loop that iterates for each episode. In each iteration $t$, the algorithm selects an action $a_t$ based on the current state $s$ of the environment, as determined by the actor-network $\pi_{\theta}(s)$. This action is executed in the environment, and the algorithm observes the resulting reward $r_t$, the new state $s_{t+1}$, and the done status $d$, which indicates whether the episode has ended.

The next step is calculating the TD target value $y$. The advantage $A$ is estimated from the TD error (difference between the TD target value $y$ and the predicted state value). The advantage is then used to update the actor-network by minimizing the loss function $L_\text{actor}(\theta)$.
and the critic-network $V_{\omega}(s)$ by minimizing the loss function $L_\text{critic}(\omega)$.

Finally, the state $s_t$ of the environment is updated to the new state $s_{t+1}$, and the algorithm repeats the loop for the next iteration. This process continues until the maximum number of episodes has been reached, or $d$ is set to $True$, at which point the A2C algorithm terminates.

\subsection{Baselines}
\label{sec:baselines}
\begin{figure*}[!htbp]
\centering
\subfloat[\centering Actor neural network for comparison with QA2C]{{\includegraphics[width=0.23\textwidth]{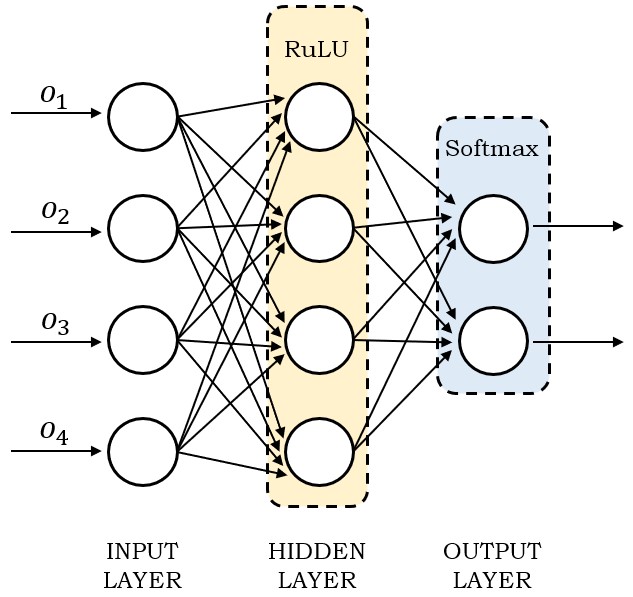} }}
\hfill
\subfloat[\centering Critic neural network for comparison with QA2C]{{\includegraphics[width=0.23\textwidth]{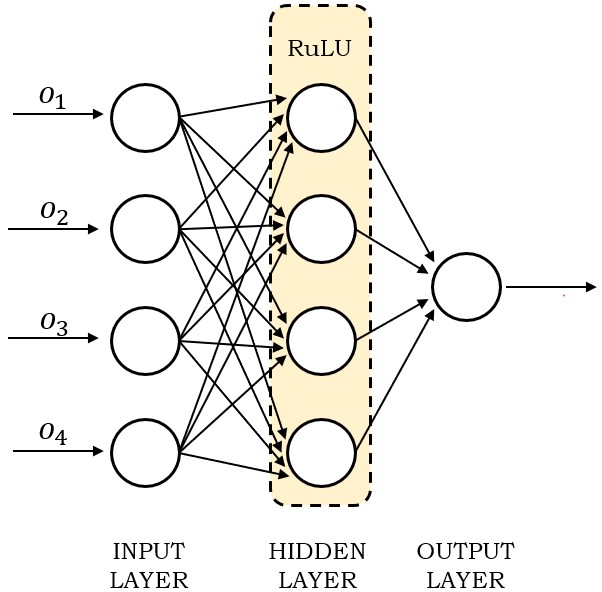} }}
\hfill
\subfloat[\centering Actor neural network for comparison with HQA2C]{{\includegraphics[width=0.23\textwidth]{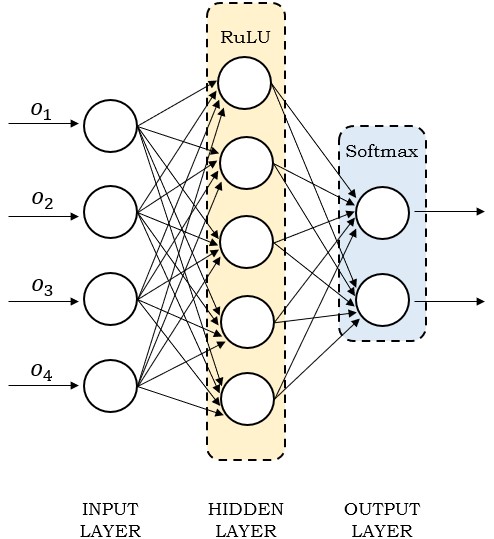} }}
\hfill
\subfloat[\centering Critic neural network for comparison with HQA2C]{{\includegraphics[width=0.23\textwidth]{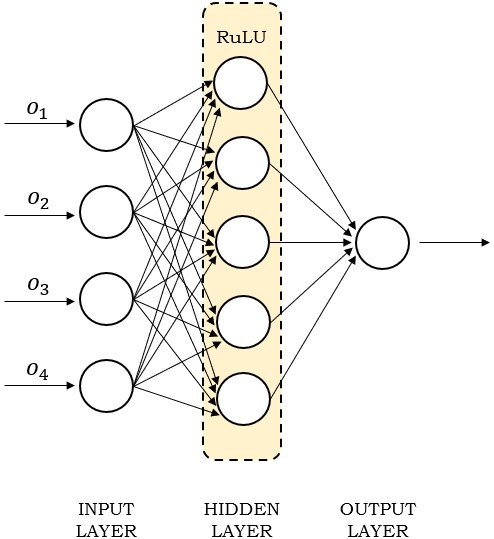} }}
\captionsetup{labelfont=bf, format=plain}
\caption{Actor and critic neural networks architectures used in A2C implementations as classical benchmarks against the QA2C and HA2C algorithms}
\label{fig:combined_nn}
\end{figure*}

In this paper, to analyze and compare the performance of quantum algorithms, we also implemented the classical A2C with different sized neural networks. The classical benchmarks were created using architectures that resemble the quantum model, leading to comparable model sizes. A2C with four hidden neurons is compared to quantum A2C, while the one with five hidden neurons is compared to hybrid A2C, described in \cref{sec:q-a2c} and \cref{sec:hq-a2c}, respectively. \cref{fig:combined_nn} shows the different neural networks.

The neural network architecture consists of an input layer, a single hidden layer, and an output layer. The input layer has four neurons that encode the 4-dimensional observation state. For the actor-network, the output layer has two neurons, while the critic network has one output neuron. We used a neural network with four neurons in the hidden layer to compare it with the VQC in the quantum A2C algorithm and five neurons compared to the VQCs with post-processing in the hybrid A2C algorithm. To ensure a fair comparison between the classical and quantum approaches, we selected these specific numbers of neurons in a hidden layer such that the number of trainable parameters in both the classical and quantum ansatz was roughly equivalent. This was done intentionally to ensure that any differences in performance between the two approaches could be attributed to their inherent differences rather than the number of trainable parameters.

For the activation function in the hidden layer, we used the Rectified Linear Unit (ReLU) activation function \cite{agarap2018deep} for both actor and critic networks. ReLU is a commonly used activation function in neural networks and helps solve the vanishing gradient problem. Additionally, in the actor-network, we used a softmax activation function \cite{BridleSoftmax} for the output layer to obtain the probability distribution.

\subsection{Quantum Advantage Actor-Critic Algorithm}
\label{sec:q-a2c}
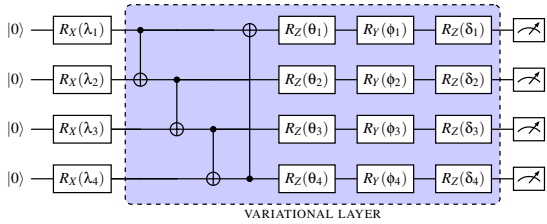
\begin{figure}[!htbp]
\begin{center}
\scalebox{0.58}{
\begin{quantikz}
\lstick{\ket{0}} & \gate{R_X(\lambda_1)} & \ctrl{1}\gategroup[4,steps=7,style={dashed,
rounded corners,fill=blue!20, inner xsep=2pt},
background,label style={label position=below,anchor=
north,yshift=-0.2cm}]{{\sc variational layer}} & \qw & \qw & \targ{} & \gate{R_Z(\theta_1)} & \gate{R_Y(\phi_1)} & \gate{R_Z(\delta_1)} & \meter{} \\
\lstick{\ket{0}} & \gate{R_X(\lambda_2)} & \targ{} & \ctrl{1} & \qw & \qw & \gate{R_Z(\theta_2)} & \gate{R_Y(\phi_2)} & \gate{R_Z(\delta_2)} & \meter{} \\
\lstick{\ket{0}} & \gate{R_X(\lambda_3)} & \qw & \targ{} & \ctrl{1} & \qw & \gate{R_Z(\theta_3)} & \gate{R_Y(\phi_3)} & \gate{R_Z(\delta_3)} & \meter{} \\
\lstick{\ket{0}} & \gate{R_X(\lambda_4)} & \qw & \qw & \targ{} & \ctrl{-3} & \gate{R_Z(\theta_4)} & \gate{R_Y(\phi_4)} & \gate{R_Z(\delta_4)} & \meter{}
\end{quantikz}
}
\end{center}
\captionsetup{labelfont=bf, format=plain}
\caption{VQC architecture utilized by QA2C and HA2C algorithms}
\label{fig:vqc_fora2c}
\end{figure}

In this work, we employ a specific VQC architecture, shown in \cref{fig:vqc_fora2c}, for both actor and critic quantum circuits in the QA2C and HA2C algorithms. As detailed in \cref{sec:vqc}, a VQC comprises three main components: an encoding layer, repeated variational layers, and measurements.

The proposed VQC utilizes $RX$ quantum gates to encode the observed state of the environment into a quantum state. For the Cart Pole problem, the observed state is four-dimensional, necessitating four qubits in both the actor and critic circuits to represent the state information.

Following the encoding process, a variational layer is applied, repeated a specific number of times $n$. In this particular implementation, we use $n = 2$ repetitions. Each variational layer consists of four CNOT gates to entangle all qubits and three single-qubit gates, $R_Z(\theta_i)$, $R_Y(\phi_i)$, and $R_Z(\delta_i)$, applied to each qubit $i$ \cite{KwakQRL_With_pennylane}. The parameters $\theta$, $\phi$, and $\delta$ are iteratively optimized using the classical optimization algorithm Adam \cite{https://doi.org/10.48550/arxiv.1412.6980}.

Lastly, each qubit's state is measured, and the measurement outcomes are utilized to determine the action for the actor and the state value for the critic in the following steps of the A2C algorithm.

\subsubsection{State Encoding}
In this work, the $RX$ gate is employed for encoding, which acts on a single qubit and performs a rotation around the $x$-axis of the Bloch sphere \cite{nielsen2010quantum}. Given that $RX$ rotations are periodic with a period of $2\pi$, different values might map to the same quantum state, leading to inaccurate predictions. To mitigate this issue, additional operations are applied to the observed state variables to ensure the parameters fall within the range of $[-\pi, \pi]$.

As shown in \cref{tab:state_variables}, the first two variables $o_1$ and $o_2$ have finite ranges, whereas the last two variables $o_3$ and $o_4$ have infinite ranges. Consequently, we establish separate normalization and transformation rules for these two groups of variables.

For the cart's position $o_1$ and the pole's angle $o_2$, which have finite ranges, the normalization procedure consists of simple scaling using their respective minimum and maximum values provided in \cref{tab:state_variables}:

\begin{equation}
\lambda_{1} = \frac{\pi}{4.8}o_1, \lambda_{2} = \frac{\pi}{0.418}o_2,
\end{equation}
where $\lambda_{1}$ and $\lambda_{2}$ denote the transformed variables that are input into the quantum circuit.

The normalization process is more intricate than simple scaling for the cart's velocity $o_3$ and the pole's angular velocity $o_4$, which have infinite ranges. To normalize these values, we first use the $\arctan$ function to map the infinite range to the finite interval $[-\pi/2, \pi/2]$ and then apply simple scaling to stretch the interval to the desired range. The process can be expressed as follows:

\begin{equation}
\lambda_{3} = 2\arctan o_3, \lambda_{4} = 2\arctan o_4
\end{equation}
where $\lambda_{3}$ and $\lambda_{4}$ represent the transformed variables used in the quantum circuit.

\subsubsection{Measurement and Action Selection of the Quantum Actor}
The actor's VQC in the QA2C algorithm is designed to convert the observed state into a quantum state and predict the optimal action to take. As described in \cref{sec:vqc}, we use a VQC comprising four qubits and two variational layers. At the end of the circuit, the qubits are measured, and the probability of measuring 0 is employed to determine the actor's action. Given that the action space in the Cart Pole environment is two-dimensional, only the measurement values of the first two qubits are considered.

To convert these values into a probability distribution, we apply the softmax function, which maps the input vector (in this case, the probabilities of the first two qubits) to a probability distribution that sums to 1 \cite{BridleSoftmax}. The softmax function is defined as:

\begin{equation}
\operatorname{softmax}(x_i) = \frac{e^{x_i}}{\sum_j e^{x_j}}
\end{equation}
where $x_i$ represents the $i$-th element of the input vector, $e$ denotes Euler's number, and the sum is taken over all elements in the input vector. After obtaining the probability distribution, the actor selects its action stochastically by randomly choosing an action from the probability distribution, based on the probabilities assigned to each action.

\subsubsection{Measurement in Quantum Critic}

In the QA2C algorithm, the quantum critic employs a VQC to estimate the value function, which serves as a measure of the quality of a given state for the agent. The quantum circuit used in this approach is detailed in \cref{sec:vqc}. As only a single value is needed for the estimation, the quantum critic measures just the first qubit and utilizes the probability of measuring 0 as the estimated state value.

\subsection{Hybrid Quantum Advantage Actor-Critic Algorithm}
\label{sec:hq-a2c}
In this section, we explore the potential of integrating VQCs and neural networks within the A2C algorithm. As discussed in \cref{sec:vqc}, neural networks can be employed in conjunction with VQCs as pre-processing and post-processing layers \cite{https://doi.org/10.48550/arxiv.2301.05096}.

In the VQC architecture depicted in \cref{fig:vqc_fora2c}, we measured at most two qubits for the actor and one qubit for the critic. With the proposed hybrid architecture, we expand the VQC by incorporating a post-processing neural network layer. This modification allows us to measure all four qubits and scale the VQC output to the desired measurement size. For the actor, we reduce the output from four values to two, corresponding to the action space of the environment. In the case of the critic, we reduce the output from four values to a single value needed for the state value function.

We refer to the proposed architecture as the Hybrid Advantage Actor-Critic (HA2C), which will be applied to the QA2C algorithm in three different approaches. The first approach entails replacing the critic network with a hybrid VQC while retaining the neural network for the actor. The second approach involves substituting the actor-network with a hybrid VQC while keeping the critic as a neural network. Finally, we replace both the actor and critic neural networks with hybrid VQCs.

\begin{figure*}[!htbp]
\centering
\subfloat[Hybrid VQC used as hybrid actor]{\includegraphics[width=0.48\linewidth]{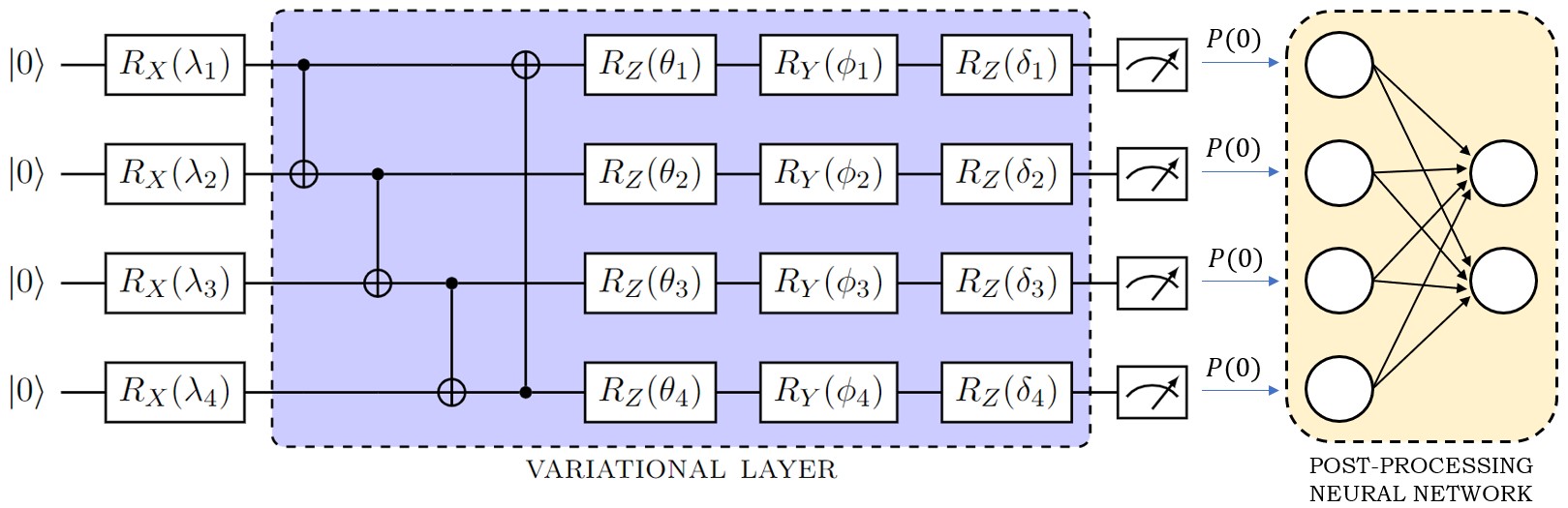}\label{fig:vqcppn_actor}}
\hfill
\subfloat[Hybrid VQC used as hybrid critic]{\includegraphics[width=0.48\linewidth]{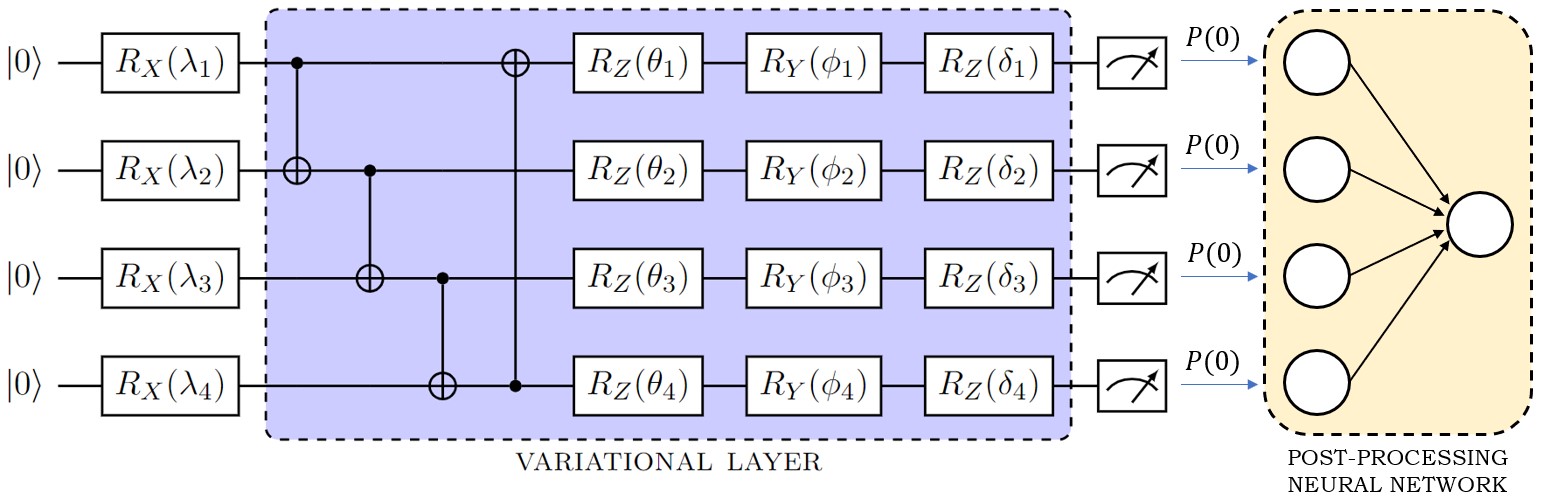}\label{fig:vqcppn_critic}}
\captionsetup{labelfont=bf, format=plain}
\caption{Hybrid VQC-based hybrid actor and critic models}
\label{fig:vqcppn_actor_critic}
\end{figure*}

\subsubsection{Hybrid Quantum Actor}
The hybrid actor in the HA2C algorithm combines a VQC, as described earlier in \cref{sec:vqc}, with a post-processing single-layer neural network, as illustrated in \cref{fig:vqcppn_actor}. The measurements obtained from the VQC are used as inputs for the post-processing layer. The neural network has four inputs and two outputs, enabling the use of measurements from all the qubits, rather than just two, as in the quantum actor in QA2C. The softmax activation function is employed at the neural network output to generate the probability distribution for the actor.

\subsubsection{Hybrid Quantum Critic}
Similarly, the hybrid critic model comprises a combination of a VQC and a post-processing neural network layer, as depicted in \cref{fig:vqcppn_critic}. The quantum circuit produces measurements that are subsequently used as input for the neural network. The critic's output is a single value, the state function, which means the neural network also has only one output. This setup allows the use of measurements from all four qubits in the VQC to estimate the state value, providing the neural network with a comprehensive set of information to process.
\section{\uppercase{Experimental Setup}} \label{sec:experimental-setup}
This paper investigates the classical A2C algorithm with two distinct implementations, each utilizing two neural networks, as described in \cref{sec:baselines} on the CartPole environment. The first implementation employs four neurons in its hidden layer, while the second uses five neurons, as shown in \cref{fig:combined_nn}. Additionally, we examine the quantum versions of A2C, which includes two main architectures: QA2C and HA2C, as explained in \cref{sec:q-a2c} and \cref{sec:hq-a2c}, respectively. Finally, for both architectures, we present three different actor-critic implementations.

In the first implementation, we replaced the neural network for the critic with a VQC in QA2C and a VQC with post-processing in HA2C, while the actor remained a neural network. We refer to this as the Advantage Actor-Quantum-Critic (A2Q) algorithm. The second implementation involved using a VQC for the actor in QA2C and a VQC with post-processing in HA2C, while the critic remained a neural network. We refer to this as the Advantage Quantum-Actor-Critic (Q2C) algorithm.

Finally, in the third implementation, we replaced both the actor and critic neural networks with a VQC in QA2C and a VQC with post-processing in HA2C. We refer to this as the Advantage Quantum-Actor-Quantum-Critic (Q2Q) algorithm.

The classical actor and critic in QA2C and HA2C used a neural network with four and five neurons in the hidden layer, respectively. Furthermore, in all experiments, the QA2C and HA2C algorithms were benchmarked against a classical A2C algorithm with four and five neurons in its hidden layer, respectively.

Classical neural networks for actor and critic were implemented using the popular deep learning library PyTorch \cite{NEURIPS2019_9015}, and we used PennyLane \cite{https://doi.org/10.48550/arxiv.1811.04968}, a widely-used quantum machine learning library, to implement VQCs. The experiments were conducted on the compute cloud provided by Leibniz-Rechenzentrum der Bayerischen Akademie der Wissenschaften (LRZ), which consists of one Intel(R) Xeon(R) Platinum 8160 CPU.

\subsection{Cart Pole}
\begin{figure}[ht]
\centering
\includegraphics[width=\linewidth]{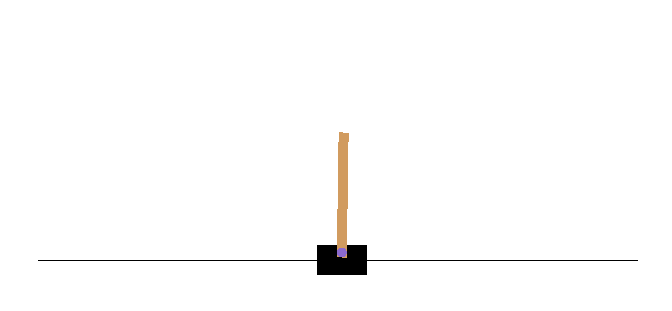}
\captionsetup{labelfont=bf, format=plain}
\caption{The Cart Pole environment \cite{brockman2016openai}}
\label{fig:cartpole}
\end{figure}

The Cart Pole testing environment provided by OpenAI \cite{brockman2016openai} serves as a popular benchmark for evaluating the performance of RL algorithms. As depicted in \cref{fig:cartpole}, the environment consists of a cart moving along a horizontal track and a pole attached to the cart by a hinge joint. The agent's objective is to balance the pole by controlling the cart's position, preventing the pole from falling over.

Several factors contribute to the Cart Pole environment's suitability as a testing environment. Firstly, its simplicity as a control problem makes it an ideal setting for testing new QRL algorithms. Secondly, it poses a relatively challenging task, as the agent must learn to balance the pole over an extended sequence of actions. Finally, the environment features a well-defined reward signal, allowing for easy evaluation and comparison of different algorithms.

The observed state consists of four state variables: the position and velocity of the cart, and the angle and angular velocity of the pole \cite{gym_cartpole}. \cref{tab:state_variables} summarizes the state variables and their respective minimum and maximum values. It is crucial to note that the Cart Pole's observed state is continuous, meaning that state variables can take any real value within a specific range. This aspect makes learning a policy more challenging since the state space is considerably larger than in environments with discrete states.

\begin{table}[ht]
\centering
\begin{tabularx}{\linewidth}{|c|C|c|c|}
\hline
Var. & Description & Min. Value & Max. Value \\ [0.5ex] 
\hline
$o_1$ & Cart Position  & -4.8 & 4.8 \\ 
\hline
$o_2$ & Pole Angle & -0.418 & 0.418 \\ 
\hline
$o_3$ & Cart Velocity & -Inf & Inf \\ 
\hline
$o_4$ & Pole Angular Velocity & -Inf & Inf \\ [1ex] 
\hline
\end{tabularx}
\captionsetup{labelfont=bf, format=plain}
\caption{Observed state variables in the Cart Pole environment}
\label{tab:state_variables}
\end{table}

The RL system's goal is to maintain the pole upright by applying actions to the cart, either moving it to the left or the right. The reward function corresponds to the time elapsed while balancing the pole. A positive reward is given at each time step as long as the pole remains upright. The system is considered to have failed if the pole falls over (pole angle not within the range of ±12°) or if the cart moves too far away from the center (greater than ±2.4), causing the episode to terminate \cite{gym_cartpole}. On the other hand, the system is deemed successful if the pole remains upright for 500 steps. This reward function encourages the RL agent to keep the pole upright for as long as possible.

\subsection{Hyperparameters and Model Size}

Hyperparameters, such as learning rates and discount factors, play a crucial role in determining the performance of RL algorithms \cite{https://doi.org/10.48550/arxiv.1709.06560}. Therefore, we conducted a small-scale hyperparameter tuning study to find a suitable learning rate $\alpha$ and discount factor $\gamma$ for the classical A2C, QA2C, and HA2C algorithms. Based on the results, we selected a learning rate $\alpha = 1\times 10^{-4}$ to be used in the Adam optimizer and a discount factor $\gamma = 0.99$ for all algorithms. All runs were executed on nodes with Intel(R) Core(TM) i5-4570 CPU @ 3.20GHz.

Our proposed VQC for the QA2C and HA2C algorithms is explained in \cref{sec:vqc} and visualized in \cref{fig:vqc_fora2c}. The VQC employs four qubits and two variational layers, with each layer consisting of three single-qubit rotations, resulting in a total of 24 quantum parameters to be optimized.
To ensure fair comparisons between the classical A2C and the quantum algorithms, we implemented the A2C algorithm with four and five neurons in the hidden layer. The exact number of parameters for all QA2C and HA2C versions are shown in \cref{tab:parameters} in (a) and (b), respectively, along with the corresponding classical A2C benchmark.

\begin{table}[ht]
\centering
\subfloat[\centering QA2C model parameters]{{\begin{tabular}{|c|c|c|c|}
\hline
& Actor & Critic & Total \\
\hline
$A2C_4$ & 30 & 25 & 55 \\
\hline
$A2Q$ & 30 & 24 & 54 \\
\hline
$Q2C$ & 24 & 25 & 49 \\
\hline
$Q2Q$ & 24 & 24 & 48 \\
\hline
\end{tabular}}}
\label{tab:qa2c_parameters}
\qquad
\subfloat[\centering HA2C model parameters]{{\begin{tabular}{|c|c|c|c|}
\hline
& Actor & Critic & Total \\
\hline
$A2C_5$ & 37 & 31 & 68 \\
\hline
$HA2Q$ & 37 & 29 & 66 \\
\hline
$HQ2C$ & 34 & 31 & 65 \\
\hline
$HQ2Q$ & 34 & 29 & 63 \\
\hline
\end{tabular}}}
\label{tab:ha2c_parameters}
\captionsetup{labelfont=bf, format=plain}
\caption{Model parameters: The total number of parameters for QA2C and HA2C algorithms is presented in tables (a) and (b) respectively, while highlighting the exact number of parameters for the actor and critic. The first row displays the number of parameters for the classical A2C algorithm, while the subsequent rows show the number of parameters for the three versions of QA2C and HA2C algorithms.}

\label{tab:parameters}
\end{table}
\section{\uppercase{Results}} \label{sec:results}

\begin{figure}[!htbp]
\centering
\includegraphics[width=\linewidth]{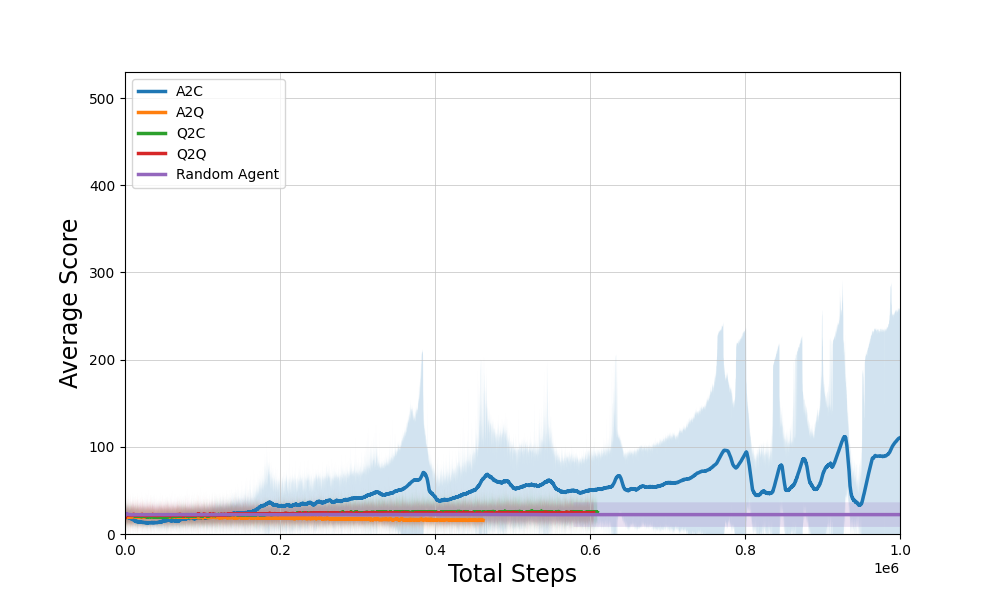}
\captionsetup{labelfont=bf, format=plain}
\caption{QA2C and classical A2C performance in Cart Pole}
\label{fig:qa2c_a2c}
\end{figure}

\begin{figure}[!htbp]
\centering
\includegraphics[width=\linewidth]{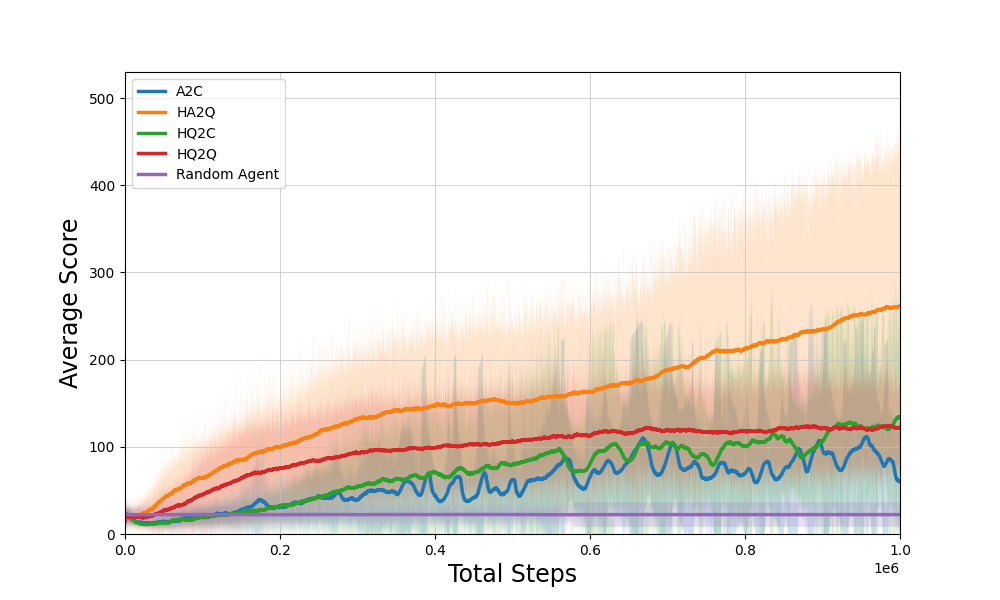}
\captionsetup{labelfont=bf, format=plain}
\caption{HA2C and classical A2C performance in Cart Pole}
\label{fig:ha2c_a2c}
\end{figure}

The first experiment aimed to compare the performance of three versions of the QA2C algorithm with the classical A2C algorithm, which had four neurons in the hidden layer. All algorithms were trained on 10 runs, each consisting of 450,000 and 1,000,000 steps in \cref{fig:qa2c_a2c} and \cref{fig:ha2c_a2c}, respectively. The $x$-axis denotes the steps, the $y$-axis shows the average reward obtained for all runs in each step, and the shaded region represents the standard deviation of the results.

The results revealed that none of the proposed quantum architectures, namely A2Q, Q2C, and Q2Q, could learn the Cart Pole environment across all runs. In contrast, classical A2C demonstrated a stable learning curve until a sudden drop at around 400,000 steps reaching a reward of 73, as shown in \cref{fig:qa2c_a2c}. However, the reward threshold for the Cart Pole environment is 475, which the classical A2C's average reward did not attain, indicating its inability to solve the environment across all runs.

The goal of the second experiment was to improve the architecture of the VQC by combining it with a post-processing neural network in order to achieve better results. \cref{fig:ha2c_a2c} shows the performance of the classical A2C with 5 neurons in the hidden layer and the three versions of the HA2C algorithm: HA2Q, HQ2C, and HQ2Q.

The aim was to train both algorithms on 10 runs, each with 1,000,000 steps in the Cart Pole environment. In \cref{fig:ha2c_a2c}, we can see an evident success for the HA2C algorithms over classical A2C, especially in HA2Q and HQ2Q. The proposed hybrid architecture learned the task in almost every run, while the classical A2C succeeded in only 5 of the 10 runs. Two additional VQCs with different single-qubit rotations were also tested in a pre-study for the QA2C and HA2C algorithms. However, the other QA2C algorithms also failed to learn.

In conclusion, the experiments demonstrated that the hybrid quantum-classical approach could achieve better results than the classical A2C algorithm with 5 neurons in the hidden layer. Specifically, the HA2C algorithms, HA2Q and HQ2Q, showed significant improvement over the classical A2C algorithm, learning the Cart Pole environment in almost every run. This highlights the potential of combining quantum and classical architectures to enhance the performance of reinforcement learning algorithms.

However, it is essential to consider the significantly longer training time for the quantum algorithms compared to the classical A2C algorithm. This performance gap emphasizes the need for further research and optimization of quantum algorithms to reduce training time and improve their applicability in real-world scenarios.

Future work could involve exploring alternative quantum architectures or optimization methods to enhance performance and reduce training time. Additionally, more complex environments and tasks could be considered to further investigate the potential of quantum-classical hybrid approaches in reinforcement learning.

\subsection{Discussion}

After conducting experiments on the Cart Pole environment, it became evident that the pure quantum A2C algorithm did not effectively learn the task. After looking at the average gradients ($-0.000056$), we concluded that this was caused by vanishing gradients. Both hybrid approaches, A2Q and Q2C suffer from the same problem using the quantum actor and quantum critic. Additionally, like with any machine learning algorithm, the performance of VQCs can be influenced by several factors, including hyperparameter settings, circuit structure, and task complexity. To improve the performance of the quantum algorithm, future work could explore techniques such as circuit ansatz design or gradient-free optimization to mitigate this issue \cite{McClean_2018,Chen_2022}. In summary, the quantum approach did not provide any significant advantage over classical methods.

Building upon that, we employ VQCs with classical post-processing to circumvent barren plateaus. This proved to be more effective in addressing the challenges of the Cart Pole environment than the classical A2C algorithm, as the hybrid quantum A2C substantially outperforms the classical A2C in learning the environment. The VQC uses a post-processing neural network, which may be crucial for enabling the VQC to learn. Notably, both HA2Q and HQ2C started to learn the task immediately, while HQ2Q learned at a slower rate.

These experiments were conducted on a quantum simulator since current quantum hardware is not widely available at the time of writing. This circumstance leads to significantly higher training times for all tested quantum approaches than the classical baseline. Thus, without access to an actual quantum device, there is currently no real benefit to the quantum approaches. Even with exclusive access to real quantum hardware, we can not say for certain if a quantum parameter trains as fast as a classical one. It is essential to recognize that the field of quantum computing and quantum machine learning is still in its early stages, with much research needed to gain a deeper understanding of the capabilities and limitations of VQCs in machine learning. As such, further exploration is necessary to unlock the full potential of this emerging technology.
\section{\uppercase{Conclusion}} \label{sec:conclusion}
In this work, we investigated how quantum computing techniques could enhance the performance of the A2C algorithm. To achieve this goal, we conducted experiments and compared the performance of three variations of the A2C algorithm: classical, quantum, and hybrid A2C. In each variation, we replaced either the actor, critic, or both with a quantum circuit, leading to a total of three different configurations. By testing these configurations, we aimed to understand the impact of each variation on the algorithm's overall performance. Furthermore, ensuring a fair comparison between the algorithms was a significant challenge in this study. Therefore, to maintain fairness, we kept the number of parameters in the algorithms roughly equal.

Our results show that the classical A2C outperforms the pure quantum A2C. To improve the quantum A2C performance, we introduced a hybrid approach integrating a VQC with a post-processing neural network layer. We tested three configurations of the hybrid algorithm and found that it substantially outperformed both the quantum and classical counterparts. This paper contributes to the growing body of evidence highlighting the potential of combining quantum computing and classical machine learning algorithms to improve reinforcement learning tasks' performance.

However, further research is still necessary. For example, tuning hyperparameters is crucial to achieving high performance in RL and QRL algorithms. In future experiments, different VQC architectures could be used to determine if the algorithm's performance can be further improved. One such technique is data re-uploading, which was not utilized in our models \cite{P_rez_Salinas_2020}. Incorporating data re-uploading may unlock additional performance gains and may be worth investigating in future studies. Another approach that could be explored is experimenting with different encoding strategies, such as amplitude encoding \cite{10.5555/3309066}. To improve convergence, one can also employ weight re-mapping to the quantum circuits \cite{kölle2023improving}. In addition, pre-processing neural networks could be explored to enhance the algorithm's performance. By testing a range of VQC architectures, we may identify those that are better suited to specific RL problems and lead to improved performance. Finally, it would be interesting to explore possible quantum variations of other state-of-the-art actor-critic algorithms, such as A3C or Proximal Policy Optimization and explore their use on different problem tasks.
\section*{\uppercase{Acknowledgements}}
This research is part of the Munich Quantum Valley, which is supported by the Bavarian state government with funds from the Hightech Agenda Bayern Plus.

\bibliographystyle{apalike}
{\small
\bibliography{main}}

\begin{thebibliography}{}

\bibitem[Agarap, 2018]{agarap2018deep}
Agarap, A.~F. (2018).
\newblock Deep learning using rectified linear units (relu).
\newblock {\em arXiv preprint arXiv:1803.08375}.

\bibitem[Andrychowicz et~al., 2021]{andrychowicz2021matters}
Andrychowicz, M., Raichuk, A., Sta{\'n}czyk, P., Orsini, M., Girgin, S., Marinier, R., Hussenot, L., Geist, M., Pietquin, O., Michalski, M., et~al. (2021).
\newblock What matters for on-policy deep actor-critic methods? a large-scale study.
\newblock In {\em International conference on learning representations}.

\bibitem[Bauer et~al., 2020]{bauer_quantum_2020}
Bauer, B., Bravyi, S., Motta, M., and Chan, G. K.-L. (2020).
\newblock Quantum {Algorithms} for {Quantum} {Chemistry} and {Quantum} {Materials} {Science}.
\newblock {\em Chemical Reviews}, 120(22):12685--12717.
\newblock Publisher: American Chemical Society.

\bibitem[Bergholm et~al., 2018]{https://doi.org/10.48550/arxiv.1811.04968}
Bergholm, V., Izaac, J., Schuld, M., Gogolin, C., Ahmed, S., Ajith, V., Alam, M.~S., Alonso-Linaje, G., AkashNarayanan, B., Asadi, A., Arrazola, J.~M., Azad, U., Banning, S., Blank, C., Bromley, T.~R., Cordier, B.~A., Ceroni, J., Delgado, A., Di~Matteo, O., Dusko, A., Garg, T., Guala, D., Hayes, A., Hill, R., Ijaz, A., Isacsson, T., Ittah, D., Jahangiri, S., Jain, P., Jiang, E., Khandelwal, A., Kottmann, K., Lang, R.~A., Lee, C., Loke, T., Lowe, A., McKiernan, K., Meyer, J.~J., Montañez-Barrera, J.~A., Moyard, R., Niu, Z., O'Riordan, L.~J., Oud, S., Panigrahi, A., Park, C.-Y., Polatajko, D., Quesada, N., Roberts, C., Sá, N., Schoch, I., Shi, B., Shu, S., Sim, S., Singh, A., Strandberg, I., Soni, J., Száva, A., Thabet, S., Vargas-Hernández, R.~A., Vincent, T., Vitucci, N., Weber, M., Wierichs, D., Wiersema, R., Willmann, M., Wong, V., Zhang, S., and Killoran, N. (2018).
\newblock Pennylane: Automatic differentiation of hybrid quantum-classical computations.

\bibitem[Biamonte et~al., 2017]{Biamonte_2017}
Biamonte, J., Wittek, P., Pancotti, N., Rebentrost, P., Wiebe, N., and Lloyd, S. (2017).
\newblock Quantum machine learning.
\newblock {\em Nature}, 549(7671):195--202.

\bibitem[Bridle, 1990]{BridleSoftmax}
Bridle, J.~S. (1990).
\newblock Probabilistic interpretation of feedforward classification network outputs, with relationships to statistical pattern recognition.
\newblock In Souli{\'e}, F.~F. and H{\'e}rault, J., editors, {\em Neurocomputing}, pages 227--236, Berlin, Heidelberg. Springer Berlin Heidelberg.

\bibitem[Brockman et~al., 2016]{brockman2016openai}
Brockman, G., Cheung, V., Pettersson, L., Schneider, J., Schulman, J., Tang, J., and Zaremba, W. (2016).
\newblock Openai gym.
\newblock {\em arXiv preprint arXiv:1606.01540}.

\bibitem[Cao et~al., 2019]{Cao_2019}
Cao, Y., Romero, J., Olson, J.~P., Degroote, M., Johnson, P.~D., Kieferov{\'{a}}, M., Kivlichan, I.~D., Menke, T., Peropadre, B., Sawaya, N. P.~D., Sim, S., Veis, L., and Aspuru-Guzik, A. (2019).
\newblock Quantum chemistry in the age of quantum computing.
\newblock {\em Chemical Reviews}, 119(19):10856--10915.

\bibitem[Cerezo et~al., 2020]{Cerezo2020VariationalQA}
Cerezo, M., Arrasmith, A., Babbush, R., Benjamin, S.~C., Endo, S., Fujii, K., McClean, J.~R., Mitarai, K., Yuan, X., Cincio, L., and Coles, P.~J. (2020).
\newblock Variational quantum algorithms.
\newblock {\em Nature Reviews Physics}, 3:625 -- 644.

\bibitem[Chen, 2023]{https://doi.org/10.48550/arxiv.2301.05096}
Chen, S. Y.-C. (2023).
\newblock Asynchronous training of quantum reinforcement learning.

\bibitem[Chen et~al., 2022]{Chen_2022}
Chen, S. Y.-C., Huang, C.-M., Hsing, C.-W., Goan, H.-S., and Kao, Y.-J. (2022).
\newblock Variational quantum reinforcement learning via evolutionary optimization.
\newblock {\em Machine Learning: Science and Technology}, 3(1):015025.

\bibitem[Chen et~al., 2021]{Chen_2021}
Chen, S. Y.-C., Huang, C.-M., Hsing, C.-W., and Kao, Y.-J. (2021).
\newblock An end-to-end trainable hybrid classical-quantum classifier.
\newblock {\em Machine Learning: Science and Technology}, 2(4):045021.

\bibitem[Chen et~al., 2019]{https://doi.org/10.48550/arxiv.1907.00397}
Chen, S. Y.-C., Yang, C.-H.~H., Qi, J., Chen, P.-Y., Ma, X., and Goan, H.-S. (2019).
\newblock Variational quantum circuits for deep reinforcement learning.

\bibitem[Chen et~al., 2020]{9144562}
Chen, S. Y.-C., Yang, C.-H.~H., Qi, J., Chen, P.-Y., Ma, X., and Goan, H.-S. (2020).
\newblock Variational quantum circuits for deep reinforcement learning.
\newblock {\em IEEE Access}, 8:141007--141024.

\bibitem[Di~Sipio et~al., 2021]{https://doi.org/10.48550/arxiv.2110.06510}
Di~Sipio, R., Huang, J.-H., Chen, S. Y.-C., Mangini, S., and Worring, M. (2021).
\newblock The dawn of quantum natural language processing.

\bibitem[Dral, 2020]{dral_quantum_2020}
Dral, P.~O. (2020).
\newblock Quantum {Chemistry} in the {Age} of {Machine} {Learning}.
\newblock {\em The Journal of Physical Chemistry Letters}, 11(6):2336--2347.
\newblock Publisher: American Chemical Society.

\bibitem[Farhi et~al., 2014]{farhi_quantum_2014}
Farhi, E., Goldstone, J., and Gutmann, S. (2014).
\newblock A {Quantum} {Approximate} {Optimization} {Algorithm}.
\newblock {\em arXiv:1411.4028 [quant-ph]}.
\newblock arXiv: 1411.4028.

\bibitem[Farhi and Harrow, 2016]{https://doi.org/10.48550/arxiv.1602.07674}
Farhi, E. and Harrow, A.~W. (2016).
\newblock Quantum supremacy through the quantum approximate optimization algorithm.

\bibitem[Gymlibrary, 2022]{gym_cartpole}
Gymlibrary, F.~F. (2022).
\newblock Cart pole - gym documentation.

\bibitem[Heimann et~al., 2022]{https://doi.org/10.48550/arxiv.2202.12180}
Heimann, D., Hohenfeld, H., Wiebe, F., and Kirchner, F. (2022).
\newblock Quantum deep reinforcement learning for robot navigation tasks.

\bibitem[Henderson et~al., 2017]{https://doi.org/10.48550/arxiv.1709.06560}
Henderson, P., Islam, R., Bachman, P., Pineau, J., Precup, D., and Meger, D. (2017).
\newblock Deep reinforcement learning that matters.

\bibitem[Homeister, 2018]{homeister2018quantum}
Homeister, M. (2018).
\newblock {\em Quantum Computing verstehen: Grundlagen -- Anwendungen -- Perspektiven}.
\newblock Computational Intelligence. Springer Fachmedien Wiesbaden.

\bibitem[Hsiao et~al., 2022]{https://doi.org/10.48550/arxiv.2203.14348}
Hsiao, J.-Y., Du, Y., Chiang, W.-Y., Hsieh, M.-H., and Goan, H.-S. (2022).
\newblock Unentangled quantum reinforcement learning agents in the openai gym.

\bibitem[Jerbi et~al., 2022]{https://doi.org/10.48550/arxiv.2212.09328}
Jerbi, S., Cornelissen, A., Ozols, M., and Dunjko, V. (2022).
\newblock Quantum policy gradient algorithms.

\bibitem[Jerbi et~al., 2021]{https://doi.org/10.48550/arxiv.2103.05577}
Jerbi, S., Gyurik, C., Marshall, S.~C., Briegel, H.~J., and Dunjko, V. (2021).
\newblock Parametrized quantum policies for reinforcement learning.

\bibitem[Kingma and Ba, 2014]{https://doi.org/10.48550/arxiv.1412.6980}
Kingma, D.~P. and Ba, J. (2014).
\newblock Adam: A method for stochastic optimization.

\bibitem[Kober et~al., 2013]{doi:10.1177/0278364913495721}
Kober, J., Bagnell, J.~A., and Peters, J. (2013).
\newblock Reinforcement learning in robotics: A survey.
\newblock {\em The International Journal of Robotics Research}, 32(11):1238--1274.

\bibitem[Konda and Tsitsiklis, 1999]{NIPS1999_6449f44a}
Konda, V. and Tsitsiklis, J. (1999).
\newblock Actor-critic algorithms.
\newblock In Solla, S., Leen, T., and M\"{u}ller, K., editors, {\em Advances in Neural Information Processing Systems}, volume~12. MIT Press.

\bibitem[Kwak et~al., 2021a]{KwakQRL_With_pennylane}
Kwak, Y., Yun, W.~J., Jung, S., Kim, J.-K., and Kim, J. (2021a).
\newblock Introduction to quantum reinforcement learning: Theory and pennylane-based implementation.
\newblock In {\em 2021 International Conference on Information and Communication Technology Convergence (ICTC)}, pages 416--420.

\bibitem[Kwak et~al., 2021b]{https://doi.org/10.48550/arxiv.2108.06849}
Kwak, Y., Yun, W.~J., Jung, S., Kim, J.-K., and Kim, J. (2021b).
\newblock Introduction to quantum reinforcement learning: Theory and pennylane-based implementation.

\bibitem[Kölle et~al., 2023]{kölle2023improving}
Kölle, M., Giovagnoli, A., Stein, J., Mansky, M.~B., Hager, J., and Linnhoff-Popien, C. (2023).
\newblock Improving convergence for quantum variational classifiers using weight re-mapping.

\bibitem[Lan, 2021]{https://doi.org/10.48550/arxiv.2112.11921}
Lan, Q. (2021).
\newblock Variational quantum soft actor-critic.

\bibitem[Lillicrap et~al., 2015]{https://doi.org/10.48550/arxiv.1509.02971}
Lillicrap, T.~P., Hunt, J.~J., Pritzel, A., Heess, N., Erez, T., Tassa, Y., Silver, D., and Wierstra, D. (2015).
\newblock Continuous control with deep reinforcement learning.

\bibitem[Lockwood and Si, 2020]{https://doi.org/10.48550/arxiv.2008.07524}
Lockwood, O. and Si, M. (2020).
\newblock Reinforcement learning with quantum variational circuits.

\bibitem[Mari et~al., 2020]{Mari_2020}
Mari, A., Bromley, T.~R., Izaac, J., Schuld, M., and Killoran, N. (2020).
\newblock Transfer learning in hybrid classical-quantum neural networks.
\newblock {\em Quantum}, 4:340.

\bibitem[McClean et~al., 2018]{McClean_2018}
McClean, J.~R., Boixo, S., Smelyanskiy, V.~N., Babbush, R., and Neven, H. (2018).
\newblock Barren plateaus in quantum neural network training landscapes.
\newblock {\em Nature Communications}, 9(1).

\bibitem[Meyer et~al., 2022a]{https://doi.org/10.48550/arxiv.2212.06663}
Meyer, N., Scherer, D.~D., Plinge, A., Mutschler, C., and Hartmann, M.~J. (2022a).
\newblock Quantum policy gradient algorithm with optimized action decoding.

\bibitem[Meyer et~al., 2022b]{https://doi.org/10.48550/arxiv.2211.03464}
Meyer, N., Ufrecht, C., Periyasamy, M., Scherer, D.~D., Plinge, A., and Mutschler, C. (2022b).
\newblock A survey on quantum reinforcement learning.

\bibitem[Mnih et~al., 2016]{https://doi.org/10.48550/arxiv.1602.01783}
Mnih, V., Badia, A.~P., Mirza, M., Graves, A., Harley, T., Lillicrap, T.~P., Silver, D., and Kavukcuoglu, K. (2016).
\newblock Asynchronous methods for deep reinforcement learning.
\newblock In {\em Proceedings of the 33rd International Conference on International Conference on Machine Learning - Volume 48}, ICML'16, page 1928–1937. JMLR.org.

\bibitem[Mnih et~al., 2015]{mnih_human-level_2015}
Mnih, V., Kavukcuoglu, K., Silver, D., Rusu, A.~A., Veness, J., Bellemare, M.~G., Graves, A., Riedmiller, M., Fidjeland, A.~K., Ostrovski, G., Petersen, S., Beattie, C., Sadik, A., Antonoglou, I., King, H., Kumaran, D., Wierstra, D., Legg, S., and Hassabis, D. (2015).
\newblock Human-level control through deep reinforcement learning.
\newblock {\em Nature}, 518(7540):529--533.
\newblock Number: 7540 Publisher: Nature Publishing Group.

\bibitem[Nielsen and Chuang, 2010]{nielsen2010quantum}
Nielsen, M. and Chuang, I. (2010).
\newblock {\em Quantum Computation and Quantum Information: 10th Anniversary Edition}.
\newblock Cambridge University Press.

\bibitem[Paszke et~al., 2019]{NEURIPS2019_9015}
Paszke, A., Gross, S., Massa, F., Lerer, A., Bradbury, J., Chanan, G., Killeen, T., Lin, Z., Gimelshein, N., Antiga, L., Desmaison, A., Kopf, A., Yang, E., DeVito, Z., Raison, M., Tejani, A., Chilamkurthy, S., Steiner, B., Fang, L., Bai, J., and Chintala, S. (2019).
\newblock Pytorch: An imperative style, high-performance deep learning library.
\newblock In {\em Advances in Neural Information Processing Systems 32}, pages 8024--8035. Curran Associates, Inc.

\bibitem[PennyLane~Team, 2022]{pennylane_variational_circuit_2022}
PennyLane~Team, X. (2022).
\newblock Variational circuit - pennylane.

\bibitem[P{\'{e}}rez-Salinas et~al., 2020]{P_rez_Salinas_2020}
P{\'{e}}rez-Salinas, A., Cervera-Lierta, A., Gil-Fuster, E., and Latorre, J.~I. (2020).
\newblock Data re-uploading for a universal quantum classifier.
\newblock {\em Quantum}, 4:226.

\bibitem[Pirandola et~al., 2020]{Pirandola_2020}
Pirandola, S., Andersen, U.~L., Banchi, L., Berta, M., Bunandar, D., Colbeck, R., Englund, D., Gehring, T., Lupo, C., Ottaviani, C., Pereira, J.~L., Razavi, M., Shaari, J.~S., Tomamichel, M., Usenko, V.~C., Vallone, G., Villoresi, P., and Wallden, P. (2020).
\newblock Advances in quantum cryptography.
\newblock {\em Advances in Optics and Photonics}, 12(4):1012.

\bibitem[Preskill, 2018]{Preskill_2018}
Preskill, J. (2018).
\newblock Quantum computing in the {NISQ} era and beyond.
\newblock {\em Quantum}, 2:79.

\bibitem[Schuld and Petruccione, 2018]{10.5555/3309066}
Schuld, M. and Petruccione, F. (2018).
\newblock {\em Supervised Learning with Quantum Computers}.
\newblock Springer Publishing Company, Incorporated, 1st edition.

\bibitem[Schulman et~al., 2017]{https://doi.org/10.48550/arxiv.1707.06347}
Schulman, J., Wolski, F., Dhariwal, P., Radford, A., and Klimov, O. (2017).
\newblock Proximal policy optimization algorithms.

\bibitem[Sequeira et~al., 2022]{https://doi.org/10.48550/arxiv.2203.10591}
Sequeira, A., Santos, L.~P., and Barbosa, L.~S. (2022).
\newblock Policy gradients using variational quantum circuits.

\bibitem[Shor, 1997]{Shor_1997}
Shor, P.~W. (1997).
\newblock Polynomial-time algorithms for prime factorization and discrete logarithms on a quantum computer.
\newblock {\em {SIAM} Journal on Computing}, 26(5):1484--1509.

\bibitem[Silver et~al., 2017]{silver_mastering_2017}
Silver, D., Schrittwieser, J., Simonyan, K., Antonoglou, I., Huang, A., Guez, A., Hubert, T., Baker, L., Lai, M., Bolton, A., Chen, Y., Lillicrap, T., Hui, F., Sifre, L., van~den Driessche, G., Graepel, T., and Hassabis, D. (2017).
\newblock Mastering the game of {Go} without human knowledge.
\newblock {\em Nature}, 550(7676):354--359.
\newblock Bandiera\_abtest: a Cg\_type: Nature Research Journals Number: 7676 Primary\_atype: Research Publisher: Nature Publishing Group Subject\_term: Computational science;Computer science;Reward Subject\_term\_id: computational-science;computer-science;reward.

\bibitem[Skolik et~al., 2022]{Skolik_2022}
Skolik, A., Jerbi, S., and Dunjko, V. (2022).
\newblock Quantum agents in the gym: a variational quantum algorithm for deep q-learning.
\newblock {\em Quantum}, 6:720.

\bibitem[Sutton and Barto, 2018]{sutton2018reinforcement}
Sutton, R. and Barto, A. (2018).
\newblock {\em Reinforcement Learning, second edition: An Introduction}.
\newblock Adaptive Computation and Machine Learning series. MIT Press.

\bibitem[Sutton et~al., 1999]{NIPS1999_464d828b}
Sutton, R.~S., McAllester, D., Singh, S., and Mansour, Y. (1999).
\newblock Policy gradient methods for reinforcement learning with function approximation.
\newblock In Solla, S., Leen, T., and M\"{u}ller, K., editors, {\em Advances in Neural Information Processing Systems}, volume~12. MIT Press.

\bibitem[You et~al., 2017]{DBLP:journals/corr/YouPWL17}
You, Y., Pan, X., Wang, Z., and Lu, C. (2017).
\newblock Virtual to real reinforcement learning for autonomous driving.
\newblock {\em CoRR}, abs/1704.03952.

\bibitem[Zhang et~al., 2020]{https://doi.org/10.48550/arxiv.2007.02151}
Zhang, J., Koppel, A., Bedi, A.~S., Szepesvari, C., and Wang, M. (2020).
\newblock Variational policy gradient method for reinforcement learning with general utilities.

\end{thebibliography}


\end{document}